\documentclass[twocolumn,showpacs,superscriptaddress]{revtex4}

\usepackage{amsmath,amssymb}
\usepackage{graphicx}
\usepackage{dcolumn}
\usepackage{bm}
\usepackage{epstopdf}
\usepackage{amstext}

\begin{document}

\title{Coherent propagation of a single photon in a lossless medium: $0\pi$ pulse formation, slow photon, storage and retrieval in multiple temporal modes }

\author{Sh. Petrosyan }
\email{shushanpet@gmail.com} \affiliation{Institute for Physical
Research, Armenian National Academy of Sciences, Ashtarak-2, 0203,
Armenia}
\author{Yu. Malakyan}
\email{yumal@ipr.sci.am}
\affiliation{Institute for Physical Research, Armenian National
Academy of Sciences, Ashtarak-2, 0203, Armenia}
\affiliation{Centre of Strong Field Physics, Yerevan State
University, 1 A. Manukian St., Yerevan 0025, Armenia}

\date{\today}

\begin{abstract}
Single-photon coherent optics represents a fundamental importance for the investigation of quantum light-matter interactions. While most work has considered the interaction in the steady-state regime, here we demonstrate that a single-photon pulse shorter than any relaxation time in a medium propagates without energy loss and is consistently transformed into a zero-area pulse. A general analytical solution is found for photon passage through a cold ensemble of $\Lambda$-type atoms confined inside a hollow core of a single-mode photonic-crystal fiber. We use the robust far off-resonant Raman scheme to control the pulse reshaping by an intense control laser beam and show that in the case of cw control field, for exact two-photon resonance, the outgoing photon displays an oscillating temporal distribution, which is the quantum counterpart of a classical field ringing, while for nonzero two-photon detuning a slow photon is produced. We demonstrate also that a train of readout control pulses coherently recalls the stored photon in many well-separated temporal modes, thus producing time-bin entangled single-photon states. Such states, which allow sharing quantum information among many users, are highly demanded for applications in long-distance quantum communication.
\end{abstract}

 \maketitle
\section{\protect\normalsize INTRODUCTION}

In classical optics, the propagation of ultrashort pulses of duration much smaller than the inverse spectral width of the absorption line is governed by the pulse-area theorem \cite{1,2}, giving us access to unusual features of the pulse propagation, which are impossible to grasp directly. In the weak field regime, the pulse area tends to zero with the propagation distance, while the pulse energy remains unchanged thus forming zero-area $(0\pi)$ pulse \cite{3,4}. These two seemingly conflicting effects simultaneously occur, if the field envelope experiences a periodic phase reversal due to alternate absorption and stimulated emission of the pulse energy that leads to a ringing structure of the output pulse. For the first time the ringing phenomenon for picosecond pulses has been experimentally demonstrated in sodium vapors \cite{5}, followed more experimental works including the pulse reshaping in resonant two-level systems \cite{6}.

Despite the fundamental importance, this problem has not yet been explored for quantum pulses, perhaps because of the limited technical abilities, in particular, the lack until recently of robust and efficient single-photon (SP) sources and detectors, as well as due to weakness of atom-photon interaction in conventional conditions. Rapid growth of quantum information applications over the past few decades has opened promising routes to circumvent these difficulties. The requirements in this area have stimulated intense research in developing truly deterministic SP sources and sensitive SP detectors \cite{7}. Furthermore, remarkable advances have been made to confine tightly the photons and a small ensemble of cold atoms inside a microscopic hollow core of a single-mode photonic-crystal fiber (HC-PCF) of a few microns in diameter \cite{8,9,10,11,12,13,14}. This architecture offers a drastic enhancement of the atom-photon interaction due to increasing the electric field amplitude of single photons. Such enhancement together with practically lossless propagation of the guided light over meter-long distances makes the HC-PCF as an ideal system to perform nonlinear optics at extremely low light levels \cite{15,16,17,18}.

In this paper we study the propagation of a short SP pulse in a cold atomic ensemble, which is assumed to be trapped inside a HC-PCF, while the incoming photon is in the fiber mode. Our research is motivated not only by the desire to understand the propagation dynamics in the quantum case, but also by the potential ability of the system to distribute the SP among well-separated temporal modes (time bins). It has been shown \cite{19} that the time-bin entanglement, in contrast to other ones, is insensitive to noises and easier to purify and, thereby, can be transferred over significantly large distances without appreciable losses. The  generation of optical pulses in distinct temporal modes has been studied experimentally in a variety of settings and under different perspectives. For example, the preparation of time-bin entangled broadband photons produced via spontaneous parametric down conversion in crystals has been studied in \cite{19,20,21}. The conversion of weak signal pulses into multi temporal modes in room-temperature vapors has been shown in EIT \cite{22} and far-off resonant Raman schemes \cite{23}. The generation of time-bin entangled photon pairs in the 1.5$\mu$m band via spontaneous four-wave mixing in a cooled fiber is reported in \cite{24}. Theoretically there exist different models to realize single photon time-bin entangled states based, for example, on the parametric interaction between two single-photon pulses in coherent atomic ensemble \cite{25}, on stimulated Raman adiabatic passage in a single quantum dot \cite{26} or in a atom-cavity system \cite{27}. Note that the time-bin entangled states are clearly well-defined provided that the temporal profiles of the modes are known, which, however, is not always available. Below we show that the lossless interaction of an ultrashort SP pulse with coherent medium is an analytically integrable model that allows a complete analysis of the propagation and controllable time-bins splitting of the SP pulse.

Specifically, our model describes a cold ensemble of $N$  $\Lambda-$atoms, which are trapped inside a HC-PCF of a small diameter $D$ and a length $L$ (see Fig.1). The atoms are initially prepared in the state $|1\rangle$ by optical pumping.  They are coupled to copropagating narrow-band quantum and control classical fields on the $|1\rangle \rightarrow |3\rangle$ and $|2\rangle \rightarrow |3\rangle$ transitions, respectively, in far off-resonant Raman configuration (Fig.1b) that makes the system immune to spontaneous losses into field modes other than the fiber mode, as well as to dephasing effects induced by other excited states. Moreover, the strong confinement of atoms in the transverse direction inside the fiber core prevents atom-wall collisions \cite{17}. Therefore, even for sub-microsecond pulses, all relaxations including the residual Doppler broadening, as well as the pulse attenuation, can be neglected thus ensuring the robust and efficient reshaping of the SP pulse.

\begin{figure}[b] \rotatebox{0}{\includegraphics* [scale =
0.50]{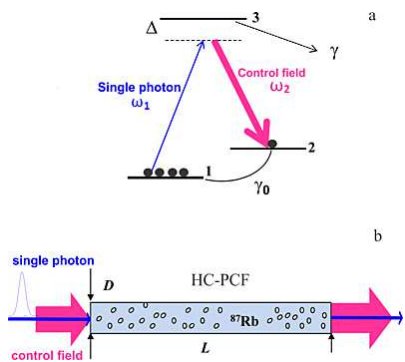}} \caption{a) Level configuration of  $\Lambda$-atoms in far off-resonant Raman scheme. The atoms are prepared in the ground state  $|1\rangle$ by optical pumping. The SP and control fields are tuned to two-photon resonance with  $|1\rangle \rightarrow |2\rangle$ transition, while one-photon detuning  $\Delta$ from the excited state  $|3\rangle$ is much larger as compared to all relaxation rates and Rabi frequencies of the fields. b) Schematic setup for SP interaction with atoms trapped in HC-FCF.}
\end{figure}

The incoming $\omega_1$-photon is converted into a Stokes photon via Raman scattering stimulated by the strong $\omega_2$-laser field and, since all the atoms are identically and strongly coupled to the fiber mode, the incident photon is stored in the medium as a symmetrically distributed single spin excitation. The probability of Stokes photon emission is less than one, so the incident photon is stored in the medium only partially, while the unstored part exits the medium without absorption. In turn, the spin wave stored in the memory is subsequently converted back by the control field into the anti-Stokes $\omega_1$-photon indistinguishable from the incident one. This conversion is clearly a direct consequence of the induced coherence between ground atomic states. Both conversion processes are highly efficient due to the fiber enhanced atom-photon interaction and multiatom collective interference effect \cite{28}. In the case of cw control beam, the process is repeated many times leading to the alternate storage and retrieval of the $\omega_1$-photon until the stored part of the incident photon is completely retrieved. As a result, the outgoing photon displays vanishing temporal oscillations with the total duration extending over many times the input pulse width. Another possibility is realized, when the control field is switched off after the photon storage and subsequent readout control pulses are applied with programmable delay in order to retrieve the stored  $\omega_1$-photon into well-separated temporal modes. In both cases the basic requirement for implementation of our scheme is the noiseless long-lived Raman memory with decoherence time exceeding the whole time of the storage and complete retrieval of the SP pulse. In recent years, by exploiting the magnetic field insensitive state and reducing the decay rates of atomic coherence, quantum memory time at single-photon level has been increased from microseconds \cite{29,30,31} to milliseconds time scale \cite{32,33,34}.

This paper is organized as follows. In the next section we derive the SP propagation law in general case of a time-dependent control field and arbitrary value of the two-photon Raman detuning. In Sec.III, we apply this solution to the case of cw control field and zero Raman detuning and show the temporal oscillations of the outgoing photon. We calculate the SP pulse area, which exponentially decreases with increasing the traveled distance. We study dispersive effects for nonzero two-photon detuning and show that the different regimes of photon propagation ranging from a slow photon to temporally oscillating photon are realized in different regions of medium dispersion. We reproduce very well the experimental results in the slow-photon regime. Then, in Sec. IV we study the SP splitting into many distinct entangled temporal modes, when a train of readout control pulses is applied after turning off the laser beam. Here we analytically obtain the profiles of temporal modes and construct the entangled state of the outgoing SP by introducing the creation operators of quantum modes satisfying the standard boson commutation relations. We also show that in our case the mode amplitudes are easily controlled by the intensities and relative phases of readout pulses. Our conclusions are summarized in Sec. V. Finally, in Appendix A we present a derivation of analytic solutions for the field operators and atomic coherence, while the pulse area is found in Appendix B.

\section{MODEL AND BASIC EQUATIONS}
We describe the quantum field by a slowly varying dimensionless operator $\hat{\mathcal{E}}(z,t)$

\begin{equation}
\label{1} E_{1}(z,t)= \sqrt{\frac{\hbar\omega_{1}}{2\varepsilon_0V}}\hat{\mathcal{E}}(z,t)
exp[i(k_{1}z-\omega_{1}t)]+H.c
\end{equation}
where $V= \pi w_a^2L$ is the volume of the medium with $w_a$ the width of the atomic transverse distribution.
The operator  $\hat{\mathcal{E}}(z,t)$ obeys the commutation relation
\begin{equation}
\label{2} [\hat{\mathcal{E}}(z,t),\hat{\mathcal{E}}^\dagger(z,t')]= \frac{L}{c}\delta(t-t')
\end{equation}
which is valid in free space $z\leq0$ and is preserved in the medium.

In the far detuned regime  $\Delta\gg\Omega_c,g,\gamma$ the upper state 3 can be adiabatically eliminated that leads to the effective Raman coupling $\hat{G}(z,t)=g \Omega^{*}_c(t)\hat{\mathcal{E}}(z , t) /\Delta$. Here $\Delta=\omega_1-\omega_{31}=\omega_2-\omega_{32}$  is the one-photon detuning, $g=\mu_{31}\sqrt{\frac{\omega_{1}}{2 \varepsilon_0\hbar V}}$ is the atom-photon coupling constant in the medium with  $\mu_{\alpha\beta}$   the dipole matrix element of the atomic  $\alpha\rightarrow\beta$   transition and $\gamma$  the upper level spontaneous decay rate. We consider the general case of time-dependent control field $E_c(t)$ described by the Rabi frequency $\Omega_c(t)=\mu_{23}E_c(t)/\hbar$.
The medium is described by atomic operators $\hat{\sigma}_{\alpha\beta}=\frac{1}{N_z} \sum \limits_{j=1}^{N_z}|\alpha\rangle_j\langle\beta|,\alpha,\beta = 1,2$,  averaged over the volume containing many atoms $N_z= \frac{N}{L} dz\gg1$ around the position $z$ \cite{35}. Here $\hat{\sigma}^{(j)}_{\alpha\beta}=|\alpha\rangle_j\langle\beta|$ is the atomic spin-flip operator in the basis of two ground states $|1\rangle$ and $|2\rangle$ for the \textit{j}- th atom.

In the dipole and rotating wave approximations the interaction Hamiltonian is given by
\begin{equation}
\begin{split}
\label{3} H=\hbar\frac{N}{L}\int_{0}^L dz[\delta_{tot}(t)\hat{\sigma}_{22}(z,t)- \\(\hat{G}(z,t)\hat{\sigma}_{21}(z,t)+h.c.)],
\end{split}
\end{equation}
where  $\delta_{tot}(t)=\delta_R+|\Omega_{c}(t)|^{2}/{\Delta}$ is the total two-photon detuning, which includes the Raman detuning $\delta_R=\omega_1-\omega_2-\omega_{21}$ and the Stark shift of the ground state 2 induced by the control field. The last terms in Eq.\eqref{3} describe the two-photon excitation of the atoms on the $1\rightarrow2$ transition.

We treat the problem in the one-dimensional (1D) approximation provided that the control beam is much wider than the SP mode and the atoms are tightly confined by the guided dipole trap with $w_a<D$  \cite{17,18}. In this case the transverse profile of the control field can be considered constant, thus reducing the propagation equations for the operator $\hat{\mathcal{E}}(z,t)$ to 1D-equation. Then the equations of the system read

\begin{equation}
\begin{split}
\label{4} (\frac{\partial}{\partial t} + c\frac{\partial}{\partial z})\hat{\mathcal{E}}(z,t)=-k\hat{\mathcal{E}}(z,t) + igN\frac{\Omega_c}{\Delta}\hat{\sigma}_{12}(z,t) + \\ +\hat{F}(z, t)
\end{split}
\end{equation}

\begin{equation}
\begin{split}
\label{5}
\frac{\partial\hat{\sigma}_{12}}{\partial t} =-(\Gamma+\gamma_0+i\delta_{tot}(t))\hat{\sigma}_{12}+i\hat{G}\hat{n}_{at}(z,t)
\end{split}
\end{equation}

\begin{equation}
\label{6}
\frac{\partial\hat{n}_{at}}{\partial t}=2i[\hat{G}^\dagger\hat{\sigma}_{12}(z,t)-\hat{G}\hat{\sigma}_{21}(z,t)]
\end{equation}

where $\hat{n}_{at}= \hat{\sigma}_{11}-\hat{\sigma}_{22}$  is the operator of population difference between the atomic ground levels, $k=\frac{ \pi\omega_{1}\mathcal{N}\mu^2_{31}\gamma }{\hbar c\Delta^2} $  is the SP linear absorption coefficient, and $\mathcal{N} =N/V$  is the uniform density of atom distribution in the HC-PCF. The first term in Eq.\eqref{5} features the optical pumping from state  $|2\rangle$ with the rate  $\Gamma=\gamma|\Omega_c|^2/2\Delta^2$, which contributes to the emission of continuum modes outside the fiber, while $\gamma_0$ is the Raman coherence damping rate (see Fig.1) caused mainly by the atom-wall collisions. The last term in Eq.\eqref{5} is responsible for generation of atomic coherence $\hat{\sigma}_{12}$.

The commutator preserving Langevin operators $\hat{F}(z,t)$ have the properties \cite{36}

\begin{equation}
\begin{split}
\label{7} \langle\hat{F}(z,t)\rangle = \langle\hat{F}(z,t)\hat{F}(z',t')\rangle =0\\
\langle\hat{F}(z,t)\hat{F}^{\dagger}(z',t')\rangle=2k\delta(z-z')\frac{L}{c}\delta(t-t')
\end{split}
\end{equation}

We require that the photon absorption and optical pumping are strongly suppressed by imposing the conditions

\begin{equation}
\label{8} kL\ll1 \text{  and  } \Gamma T\ll 1
\end{equation}
where the interaction time $T$ in our case is the SP pulse duration. Nonetheless, we will take $\Gamma$  into account in calculation of the pulse area in order to avoid the emergent mathematical uncertainties, while, in other cases, this term can be safely ignored. As is seen from Eqs.(7), in the absence of photon losses the noise operators $\hat{F}(z,t)$ give no contribution. Below we will neglect also $\gamma_0$, and the residual Doppler shift $\sim(k_1-k_2)u$, which are much smaller as compared to other relaxation rates for the atoms trapped in a hollow fiber with the most probable longitudinal velocity $u\sim 1m/s$ [14,17]. With these approximations, the width of two-photon transition line is determined solely by the optical pumping rate $\Gamma$, as is seen from Eq.(5), and, hence, the second inequality in (8) corresponds to the sharp-line limit, when the SP pulse's spectral width $\Delta\omega_1\sim T^{-1}$ is much broader than the atomic transition linewidth. This indicates that the absorption of the photon occurs only in a very narrow spectral region around the two-photon resonance and is negligible, therefore the SP pulse reshaping predicted in the present paper is due to dispersion effects. We give in the next section a clear insight into the effect of two-photon induced dispersion on the group velocity of the SP pulse.

We now outline the solution of Eqs. \eqref{4},\eqref{5},\eqref{6} in the weak-field approximation assuming that the atomic state  $|2\rangle$ is faintly populated due to the interaction of multi-atomic ensemble with weak single photon field, so that we treat the atomic equations perturbatively in the small parameter $g\mathcal{\hat{E}}$. By taking $\hat{n}_{at}=\hat{I}$ in Eqs.\eqref{5},\eqref{6}, this leads to the linearized system of Heisenberg equations, which in the wave variables $z$ and  $\tau=t-z/c$ take the form

\begin{equation}
\label{9}
   c \frac{\partial}{\partial z}\hat{\mathcal{E}}(z,\tau)=igN \frac{\Omega_c(\tau)}{\Delta}\hat{\sigma}_{12}(z,\tau)
\end{equation}
\begin{equation}
\label{10}
   \frac{\partial\hat{\sigma}_{12}}{\partial t}=-i\delta_{tot}(\tau)\hat{\sigma}_{12}+ ig\frac{\Omega_c^*(\tau)}{\Delta}\hat{\mathcal{E}}(z,\tau)
\end{equation}
The formal analytical solution of Eqs.(9,10) is found in Appendix A in general case of arbitrary time dependence of the control field and two-photon detuning $\delta_{tot}(t)$. In what follows, we confine ourselves to the Fourier-transform control field without phase fluctuations. In that case the Rabi frequency  $\Omega_c(\tau)$ can be considered real. Then we find from Eq.(A.7)

\begin{equation}
\begin{split}
\label{11} \hat{\mathcal{E}}(z,\tau)=\hat{\mathcal{E}}(\tau)-2qz\frac{\Omega_c(\tau)}{\Delta^2}\int_{-\infty}^{\tau}d\tau' \\
\hat{\mathcal{E}}(\tau')\Omega_c(\tau')\frac{J_1(\psi)}{\psi(\tau, \tau')}\text{exp}[-i\int_{\tau'}^{\tau}\delta_{tot}(x)dx]
\end{split}
\end{equation}
where $\mathcal{\hat{E}}(\tau)=\mathcal{\hat{E}}(0,\tau), \psi(\tau, \tau')=2 \sqrt{qz\int_{\tau'}^{\tau}\frac{\Omega^2_c(x)}{\Delta^2}dx}$ and $q=\frac{1}{2}\alpha\gamma$ with $\alpha=\frac{4 \pi\omega_1\mu_{31}^2 \mathcal{N}}{\hbar c \gamma}=\frac{2g^2N}{c\gamma}$  the resonant absorption coefficient on the transition $1\rightarrow3$. By combining Eqs. \eqref{9}, \eqref{10}  one obtains

\begin{equation}
\begin{split}
\label{12} \hat{\sigma}_{12}(z,\tau)=i\frac{g}{\Delta}\int_{-\infty}^{\tau}d\tau' \Omega_c(\tau')\hat{\mathcal{E}}(\tau') \\
           [2\frac{J_1(\psi)}{\psi(\tau,\tau')} -J_2(\psi)]\text{exp}[-i\int_{\tau'}^{\tau}\delta_{tot}(x)dx]
\end{split}
\end{equation}

Using the known properties of Bessel functions, one can easily verify that the solution \eqref{11} for operator $\mathcal{\hat{E}}(z,t)$ satisfies the commutation relation \eqref{2}. Also, it is seen that the atomic coherence $\hat{\sigma}_{12}$ is purely imaginary that provides the attenuation or gain of the quantum field depending on the sign of the mean value  $\langle\text{Im}\sigma_{12}\rangle$, as it follows from Eq. \eqref{9}.

The Eqs. \eqref{11} and \eqref{12} are our central results, which will be applied in next sections for two cases of formation of a single-photon zero-area pulse and conversion of the stored atomic spin excitation into $\omega_1$-photon in distinct temporal modes by applying a train of readout pulses.

\section{\protect\normalsize SINGLE-PHOTON 0$\pi$  PULSE}
In this section, we consider the case of cw control field with $\Omega_c(\tau)\equiv\Omega_0$. We assume also that the total two-photon detuning is small compared to the spectral width of the SP field: $\delta_{tot}\ll \Delta\omega_1$ or $\delta_{tot}T\ll 1$. This condition can be fulfilled by taking $\delta_{tot}=0$, where the constant Stark shift  $\Omega_c^2/\Delta$ is incorporated into the frequency of the control field. The opposite case $\delta_{tot}T\gg 1$ will be discussed in Section V.

We are interested in the evolution of the input state  $|\Psi_{in}\rangle$  representing a single-photon wave packet with given temporal profile $f(t)$ at $z=0$.

\begin{equation}
\label{13}
\langle0\mid\hat{\mathcal{E}}(0,\tau)\mid\Psi_{in}\rangle=f(\tau)
\end{equation}

which is normalized as $\frac{c}{L}\int|f(t)|^2dt=1$ indicating that the initial number of photons is one (see below). Then from Eqs. \eqref{11} and \eqref{13} the SP field amplitude (or the photon wave function) $\Phi(z,\tau)$ and pulse intensity at any distance in the region $0\leq z\leq L$ take the form
\begin{equation}
\begin{split}
\label{14}
\Phi(z,\tau) = \langle0\mid\hat{\mathcal{E}}(z,\tau)\mid\Psi_{in}\rangle=f(\tau)\\
-2qz\int_{-\infty}^{\tau}d\tau' f(\tau') \frac{J_1(\psi_0(\tau, \tau'))}{\psi_0(\tau, \tau')}
\end{split}
\end{equation}

\begin{equation}
\begin{split}
\label{15}
I(z,t)= \langle\Psi_{in}|\hat{\mathcal{E}^\dagger}(z,\tau)\hat{\mathcal{E}}(z,\tau)|\Psi_{in}\rangle=\\ |\Phi(z,\tau)|^2
\end{split}
\end{equation}

where  $\psi_0(\tau,\tau')=2\sqrt{qz\frac{\Omega_0^2}{\Delta^2}(\tau-\tau')}$.
For the atomic coherence we obtain from Eq.\eqref{12}
\begin{equation}
\begin{split}
\label{16}
\rho_{21}(z,t)=\langle0| \hat{\sigma}_{12}(z,\tau)|\Psi_{in}\rangle =\\
 \frac{ig\Omega_0}{\Delta}\int_{-\infty}^{\tau}d\tau' f(\tau')[2\frac{J_1(\psi_0)}{\psi_0}-J_2(\psi_0)]
\end{split}
\end{equation}

It is seen that at small distances $z\sim 0$, where $\psi_0\rightarrow 0$ and $J_1(\psi_0)\rightarrow \psi_0/2$ and $J_2(\psi_0)\rightarrow 0$, the atomic coherence becomes proportional to the initial pulse area $\theta(0)=g\int_{-\infty}^{\infty}f(\tau)d\tau$.
At any distance, we define the SP pulse area as
\begin{equation}
\label{17}
\theta(z)=\lim\limits_{t\rightarrow\infty}g\int_{-\infty}^{\tau}\Phi(z,t')dt
\end{equation}
which, multiplied by  $\Omega_0/\Delta$, determines the total angle that the atomic state vector rotates around the electric field of the SP pulse. Hereafter, for simplicity,  $\Phi(z,t)$ is taken real. The solution for $\theta(z)$ is found in Appendix B by integrating the Eq.\eqref{14} over the time that yields

\begin{equation}
\label{18}
\theta(z)=\theta(0)\text{exp}(-\alpha z)
\end{equation}

Equation \eqref{18} is the area theorem for the SP pulse, which has the same form as in the classical case [1], but with different decay constant $\alpha$. If the latter is a few hundreds $cm^{-1}$, the pulse area vanishes at sub-millimeter distances, which means that the single-photon $0\pi$  pulse is formed practically almost immediately after the SP pulse has entered the medium. Meanwhile, the pulse energy remains finite. To show this we introduce the operator for number of photons that pass each point on the $z$ axis over the whole of time

\begin{equation}
\label{19}
\hat{n}(z)=\frac{c}{L}\int dt\hat{\mathcal{E}}^\dagger(z,t)\hat{\mathcal{E}}(z,t)
\end{equation}

and using Eqs.\eqref{9}, \eqref{10} we derive the equation for $n(z)=\langle\Psi_{in}|\hat{n}(z)|\Psi_{in}\rangle$ in the form

\begin{equation}
\label{20}
\frac{\partial n}{\partial z}=-\frac{N}{L}\langle\Psi_{in}|\hat{\sigma}_{22}(z,\infty)|\Psi_{in}\rangle=-\frac{N\Omega_0^2\theta^2(z)}{L\Delta^2}
\end{equation}
where the population in the excited state 2 at large times is connected to the pulse area by (see Appendix B)
\begin{equation}
\label{21}
\rho_{22}(z)=\langle\Psi_{in}|\hat{\sigma}_{22}(z,\infty)|\Psi_{in}\rangle=\Omega_0^2\theta^2(z)/\Delta^2
\end{equation}
This important relation shows that the photon capture occurs only at small distances $z\leq\alpha^{-1}$, being furthermore greatly suppressed by the factor $\Omega_0^2/\Delta^2$. For an initial Gaussian pulse with $\theta(0)=g\sqrt\pi T$, the total photon losses are found from Eqs.\eqref{20} and \eqref{18} as

\begin{equation}
\label{22}
n(0)-n(L)=\frac{cT}{L}\frac{\gamma T\Omega_0^2}{\Delta^2}
\end{equation}

\begin{figure}[b] \rotatebox{0}{\includegraphics* [scale =
0.7]{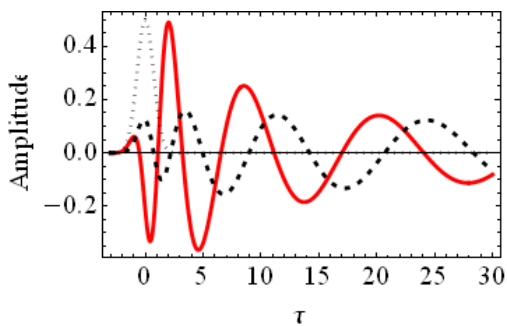}} \caption{(color online) The amplitude $\Phi(L,\tau)$ of the SP output pulse (red solid) and atomic coherence $Im\rho_{21}(L,\tau)$(black dashed) as functions of $\tau$ (in units of SP pulse duration $T=200ns$) for  $\Delta=20\gamma , \Omega_0/\Delta=0.1$ and $L=3cm$. The input pulse is shown by dotted line, scaled by factor 0.5.}
\end{figure}

\begin{figure}[b] \rotatebox{0}{\includegraphics* [scale =
0.4]{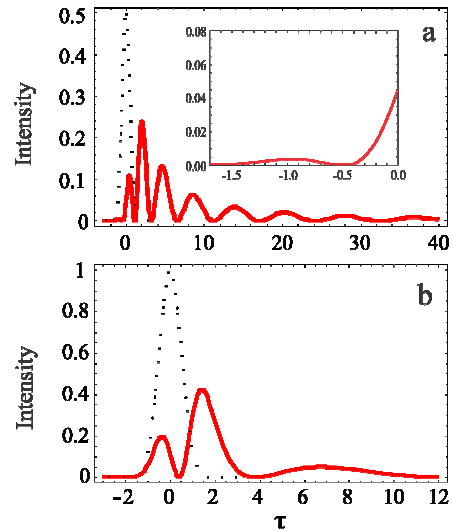}} \caption{ (color online).The output intensity of SP pulse (red solid) as a function of $\tau$ (in units of $T$) at $z= \emph{L}$ (a) and $z=0.25\emph{L}$ (b). The input pulse of duration 200ns is shown by dotted line, which in the case (a) is scaled by 0.5. The inset in this case shows an unstored part of the input SP pulse on its leading edge. For the rest of the parameters see the text.}
\end{figure}

where $n(0)=\langle\Psi_{in}|\hat{n}(0)|\Psi_{in}\rangle=\frac{c}{L}\int|f(t)|^2dt=1$ is the initial number of photons.  From this equation we recognize that in the weak-field approximation the photon losses do not depend on the number of atoms, but instead, they are proportional to the ratio of the pulse length $L_p= cT$ to the length $L$ of the medium; the larger this ratio, the greater the photon losses. The second factor in Eq.\eqref{22}, which describes the spontaneous losses due to optical pumping during the time $T$, must be small according to Eq.\eqref{8}. Thus, the required suppression of photon losses can be achieved by properly adjusting the system parameters, primarily the control field intensity and detuning. This fact is compatible with zero pulse-area in Eq.\eqref{18} only if, as mentioned previously, the photon wave function $\Phi(z,\tau')$ in Eq.(17) is an alternating function of time.

For numerical calculations we consider cold $^{87}$Rb atoms with the ground states $5S_{1/2}(F=1)$ and $5S_{1/2}(F=2)$ and excited state $5P_{1/2}(F=2)$ as the atomic states 1, 2, and 3 in Fig. 1, respectively. We show in Fig.2 the field amplitude of output photon $\Phi(L,\tau)$ and atomic coherence $\rho_{12}=\langle0|\hat{\sigma}_{12}|\psi_{in}\rangle $ as a function of time using realistic parameters to fulfill all necessary conditions minimizing the photon losses:
number of atoms confined in a hollow-core fiber of the length $L\sim 3cm$ with radial distribution width  $w_a\sim2 \mu m$ is about $N\sim10^4$ \cite{17}, the fields are tuned away from the one-photon resonance by $\Delta=20\gamma$, while the Rabi frequency of the control field is taken  $\Omega_0/\Delta=0.1$, which corresponds to the control field intensity $1mW/cm^2$. The input SP pulse of wavelength  $\lambda\sim 0.8\mu m$ has a Gaussian profile $f(t)\sim \text{exp}(-t^2/T^2)$ with duration $T\sim200ns$.

It is seen that the SP field amplitude periodically passes zero-value points which correspond to the SP storage-retrieval transitions. Numerically the pulse area in the depicted region $-2<\tau/T<30$ is about $g \int_{-\infty}^\infty\Phi(z,\tau)d\tau\sim -0.1$, while for the initial pulse it is $\sim1.2$.  At these points, according to $d\rho_{12}/dt\sim\Phi(z,t)$, the atomic coherence becomes maximal.

Correspondingly, the SP intensity given by Eq.(15) displays temporal oscillations (Fig.3), which resemble the classical field ringing \cite{5}. However, in the quantum case these oscillations represent the arrival time distribution of photons, which are recorded with single-photon detectors. The classical ringing emerges out of these random quantum events as a result of a large number of single photon measurements. Ideally, a coherent field ringing pattern can be identically reconstructed by using a train of indistinguishable single-photon pulses. A deterministic source of such photons has been recently proposed in our work \cite{37}.

Two important consequences come out of the Eq.\eqref{15} and Fig.3. First, the SP intensity $I(z,\tau)$ depends on the product $z\Omega_0^2$ and thereby, invariant under the transformation

\begin{equation}
\label{23}
z\rightarrow z/a, \Omega_0\rightarrow \sqrt a\Omega_0
\end{equation}
where $a$  is an arbitrary real constant. This allows one to follow the behavior of the SP pulse at any point $z\leq L$ inside the sample. Indeed, for given $L$ and control field Rabi frequency $\Omega_0$, the pulse intensity, for example, at  $z=L/a$ is the same as the output intensity at $z=L$, but with reduced Rabi frequency  $\Omega_0/\sqrt a$. The case of $a=4$ is shown in Fig.3b.

\begin{figure}[b] \rotatebox{0}{\includegraphics* [scale =
0.65]{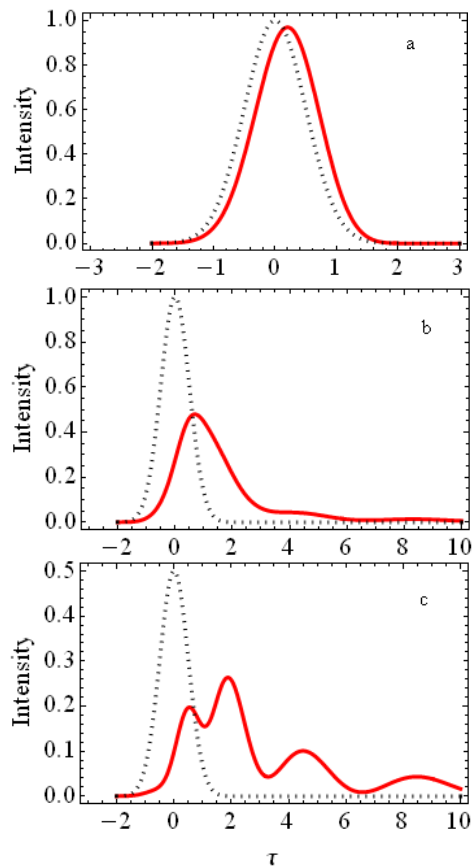}} \caption{(color online) Output photon intensity as a function of $\tau$ (in units of $T$) for three values of two-photon detuning: $\delta_{tot}T=5$ (a), 2 (b) and 1 (c). In the case (a) the slow photon regime is realized. The input pulse is shown by dotted line, scaled in the case (c) by factor 0.5. The rest parameters are the same as in Fig.2.}
\end{figure}

Secondly, if the control field is strong enough, the incoming SP pulse is almost completely stored, and only a small portion of the photon escapes the medium, which is observed in Fig.3a as a tiny spike on the leading edge of the initial pulse. The stored part of the SP pulse is then converted back via multiple retrievals thus forming a sequence of damped temporal peaks. The duration of these oscillations increases with increasing of the control field Rabi frequency and can be roughly estimated as
\begin{equation}
\label{24}
T_{out}\sim 20(qLT\frac{\Omega_0^2}{\Delta^2})^{1/3}T
\end{equation}

For the above parameters $T_{out}\sim10 T$. Meanwhile, beginning from $z\geq\alpha^{-1}$  the atomic population is not accumulated in the excited state 2, as follows from Eq.\eqref{21}, indicating that the energy transferred to this state is eventually converted back into the $\omega_1$-photon, thus conserving the photon number. But, a minor part of photon remains trapped in the excited state at small distances.  The total number of photons at the entrance and exit of the medium is determined by the areas of the corresponding peaks in Fig.3. For example, in Fig.3a, the photon losses for chosen parameters are only $\sim8\%$, which can be further lowered by using shorter SP pulses (see Eq.\eqref{22}).

Finally, we demonstrate the different regimes of the SP pulse propagation in dependence on the medium dispersion and discuss the role of dispersive effects in the pulse reshaping. In Fig.4, the output photon intensity is calculated for three values of two-photon detuning $\delta_{tot}$ using Eq.(11) for constant $\Omega_c(\tau)$. For the large value of $\delta_{tot}T=5$ (Fig.4a), the ordinary dispersive interaction occurs without absorption or storage of the photon and, thereby, the group velocity of the SP pulse is reduced while preserving its shape. As a result, a slow photon regime is realized with pulse delay $\sim 0.2T$. Harris predicted this effect for classical probe light as early as 1994 \cite{38}, which has been then confirmed experimentally in solid hydrogen \cite{39,40}. For smaller detuning $\delta_{tot}T=2$ (Fig.4b), the atomic two-photon resonance transition is in the wings of the SP pulse spectrum and the photon is partially stored. The corresponding spectral components experience steeper dispersion, resulting in that the SP pulse evolves to a two-peak structure, where the group delay of the second peak is significantly larger than the SP pulse width. We believe just this effect for a classical Stokes field has been observed in SRS in solid hydrogen \cite{40}, although the authors give another interpretation. The dispersion effects are enhanced with $\delta_{tot}$ decreasing, which leads to a multi-peak structure of the output pulse (Fig.4c) and eventually, when $\delta_{tot}=0$, the pulse is split into separate subpulses depicted in Fig.3a. Let us note once more that in all cases the real absorption of the photon is negligible and has been ignored. In the last case, taking into account that $f(\tau')$ in Eq.(14) is localized around $\tau'=0$ and, hence, from Eqs.(14,15) the pulse intensity can be presented as $I(z,t)\sim J_1^2(\psi_0(\tau, 0))$, the relative delay between the subpulses for a given $z$ can be found from the values of $\psi_0^{\text{extr}}=2\sqrt{qz\frac{\Omega_0^2}{\Delta^2}\tau^{\text{extr}}}$, which are extremum points of the Bessel function $J_1^2(\psi_0(\tau, 0))$. Since $\tau^{\text{extr}}\sim 1/z$, the delay time decreases with $z$ increasing, as is seen from Figs.3a and 3b.

\section{RETRIEVAL OF STORED PHOTON INTO MULTI-TIME BINS}

In this section, we discuss the SP splitting into many well-separated entangled temporal modes. At first the SP is stored in an atomic medium for a long time via a standard technique \cite {35} by turning off the control field just as the SP pulse enters the medium. After some delay, the stored atomic spin wave is converted back into original  $\omega_1$-photon by applying a train of readout control pulses of $\omega_2$ frequency with properly adjusted amplitudes and phases to produce the desired states of regenerated  $\omega_1$-photon. A similar approach has been used previously for weak light pulses in room temperature alkali vapor \cite{23}, but its application to a single-photon case may require serious modifications of the setting.

The control field $\Omega_c(\tau)$ is now chosen as

\begin{figure}[b] \rotatebox{0}{\includegraphics* [scale =
0.4]{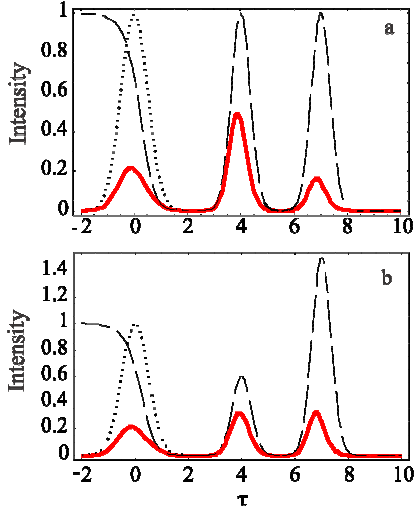}} \caption{ (color online) Twofold retrieval of stored $\omega_1$-photon (red-solid line) by two readout control pulses (thin dashed line) localized at  $\tau_1=4T$ and  $\tau_2=7T$ with durations $T_1=T_2= T / \sqrt{2}$, respectively, and having Rabi frequencies $\Omega_1=\Omega_2=\Omega_0$ in (a) and  $\Omega_1= 0.08\Omega_0, \Omega_2=1.2\Omega_0$ in (b). In both cases, the Rabi frequency of initial control field $\Omega_0=0.05\Delta$  with falling time $T_0=1.5T$. The left peaks on red lines represent the transmitted part of incident SP pulse (dotted) and obviously is the same in both cases. For the rest parameters see the text.}
\end{figure}

\begin{figure}[b] \rotatebox{0}{\includegraphics* [scale =
0.7]{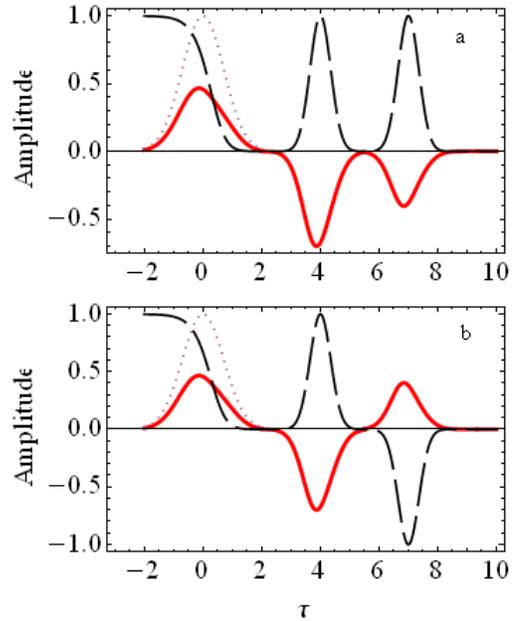}} \caption{(color online) Amplitudes of outgoing  $\omega_1$-photon (red-solid line) for the case of Fig.4a. The two readout control pulses (dashed line)  have the same (a) and opposite (b) phases.  The incident SP field amplitude is shown by dotted lines. }
\end{figure}

\begin{equation}
\label{25}
\Omega_c(\tau)=\Omega_0f_0(\tau-\tau_0)+\sum\limits_{i=1}^{J}\Omega_if_i(\tau-\tau_i)
\end{equation}
where the $J$ well-separated readout pulses are localized at time moments  $\tau_J >\tau_{J-1}... >\tau_1>\tau_0$  with the relative delay between them much larger than their lengths $T_i$. In the first term of Eq.(25), the function  $f_0(\tau -\tau_0)$ implements the switching of the control field from its constant value  $\Omega_0$ to zero with the rate $T_0^{-1}$, where $T_0\sim T \sim T_i$. In what follows, we take $f_0(\tau -\tau_0)$ in the form $f_0(\tau -\tau_0)=\frac{1}{2}[\text{Tanh}(-\frac{\tau-\tau_0}{T_0})+1]$.

To find the mode amplitudes we use the solution for the field operator Eq.(11) with time-dependent control field and $\delta_{tot}=0$, which is now satisfied, if $\delta_R=0$ and $\Omega^2_{j}T_{j}/\Delta<<1$. From Eqs.(11) and (25), the SP wave function takes the form
\begin{equation}
\label{26}
\Phi(z,\tau) = \langle0\mid\hat{\mathcal{E}}(z,\tau)\mid\Psi_{in}\rangle=\sum_{i=0}^J\Phi_i(z,\tau)
\end{equation}
where the profile of the $i$-th temporal mode is given by
\begin{equation}
\label{27}
\Phi_i(z,\tau)=f(\tau)\delta_{i0}-2qz\frac{\Omega_0\Omega_i f_i(\tau-\tau_i)}{\Delta^2}I(z,\tau)
\end{equation}
with
\begin{equation}
\label{28}
I(z,\tau)=\int_{-\infty}^\tau d\tau'f(\tau')f_0(\tau'-\tau_0) \frac{J_1(\psi(\tau,\tau'))}{\psi(\tau,\tau')}
\end{equation}
Here the orthogonality of $f_i(\tau-\tau_i), i\neq0,$ to the input wave function $f(\tau)$ has been used.

It is seen from Eq.(27) that the profiles of all modes are proportional to the same function $I(z,\tau)$, in which the product $f(\tau')\Omega_0(\tau')$ in the integrand is responsible for the storage of the initial photon at $\tau'\sim0$, in the time localization of $f(\tau)$, while the reduction of the stored part of the SP after each readout control pulse is given by the factor $J_1(\psi)/\psi$ . The function $\Omega_i f_i(\tau-\tau_i)$ in the second term in Eq.(27) describes the retrieving of $\omega_1$-photon in nonoverlapping temporal modes with phases of corresponding readout control pulses, thus providing the phase coherence across all time bins.
In Fig.4 we present the output $\omega_1$-pulse for two sets of control readout pulses. In the first case, the amplitudes of two control pulses are the same (Fig.4a), while in the second one they are redesigned such that two temporal modes of retrieved  $\omega_1$-photon have equal intensities (Fig.4b). In Figs.5 we show that the temporal mode amplitudes of the outgoing photon depend also on the relative phase of readout control pulses that makes it possible to produce the time-bin entangled photonic states with alternating phase.

Then, we construct the single-photon output state in terms of quantized temporal modes. For the two modes this technique has been previously developed in our work [25]. To generalize it to the multi-mode case we proceed from the fact that this state is the eigenstate of the photon number operator (19): $\hat{n}(z)|\Psi_{out}\rangle=|\Psi_{out}\rangle$, associated with the wave function (26). Using the commutation relation (2), this yields
\begin{equation}
\label{29}
\mid \Psi_{out}\rangle=\frac{c}{L}\int dt \Phi(L,t)\hat{\mathcal E}^{\dagger}(L,t)\mid0\rangle
\end{equation}
with normalization
\begin{equation}
\label{30}
\frac{c}{L}\int dt \mid \Phi(L,t)\mid^2=1
\end{equation}
Each mode function $\Phi_{i}(L,t)$ is assigned a creation operator defined as
\begin{equation}
\label{31}
\hat c_{i}^{\dagger}=N_{i}^{1/2}\int dt
\Phi_{i}(L,t)\hat{\mathcal E}^{\dagger}(L,t)
\end{equation}
where the normalization constant is
\begin{equation}
\label{32}
N_{i}=\frac {c}{L}(\int dt \mid\Phi_{i}(L,t)\mid^2)^{-1}
\end{equation}
These operators create the single-photon states in the usual way by operation on the vacuum: $\hat c_{i}^{\dagger}\mid0\rangle=\mid 1\rangle_{i}$ and have the standard boson commutation relations $[\hat c_{i},\hat c_{j}^{\dagger}]=\delta_{ij}$. Note that this definition of quantum temporal modes is only useful, if one can perform the local measurements on the modes such that they are spacelike separated, so that the requirement for the modes to be well separated is important.

Taking into account that the modes, which are not occupied by the photons, are not included in the measurements, the total vacuum may be reduced to the product $\mid0\rangle=\prod_{i=0}^{J}\mid0\rangle_{i}$ and, hence, from Eqs.(26) and (29), the output state of the photon can be written as an entangled state of $J+1$ temporal modes
\begin{equation}
\label{33}
\mid \Psi_{out}\rangle=\sum_{i=0}^J r_{i}\hat c_{i}^{\dagger}\mid0\rangle=\sum_{i=0}^J r_{i}\mid 1\rangle_{i}\prod_{j\neq i}^{J}\mid0\rangle_{j}
\end{equation}
where $r_{i}=\sqrt{\frac {c}{L}\int dt \mid \Phi_{i}(L,t)\mid^2}$. Accordingly, the number of photons in each mode is $n_{i}=r_{i}^2$. Due to the orthogonality of mode functions $\Phi_{i}(L,t)$ we have from Eq.(30)
\begin{equation}
\label{34}
\sum_{i=0}^J n_{i}=1
\end{equation}
In Figs.(4,5), this condition is fulfilled even for two control readout pulses, where the number of photons in the modes is determined by the areas of the corresponding peaks.

The state (33) is a pure entangled state with mode phases equal to those of the corresponding mode functions $\Phi_{i}$, as follows from Eq.(31). In bimodal case of $J=1$, the degree of entanglement and nonlocal correlations have been comprehensively studied in [25]. Meanwhile, the quantifying of multimode entanglement for $J>1$ is not still available, as the current understanding of multipartite entanglement is very vague due to the lack of a theory. However, we expect the future study of these issues may be based on Eq.(33) obtained for the first time in terms of quantized modes with known spatio-temporal profiles.

\section{CONCLUSIONS}

In this paper we have raised the question how a single photon travels in a lossless medium and how it can be stored for a long time and retrieved on demand in a desired state. To answer these questions, we have suggested a far off-resonant Raman scheme based on a cold atoms trapped in a hollow-core fiber of a small diameter. The collective atomic effects and tight transverse confinement of the atoms essentially enhance the atom-photon interaction, while the photon losses are strongly suppressed employing highly detuned Raman transitions. Our analytical results for the pulse area and field intensity demonstrate a quantitative understanding of the SP travel as continuously alternating storage and coherent retrievals of the photon leading to temporal oscillations of photon distribution, which are the quantum counterpart of a classical field ringing. We have shown the sole responsibility of dispersive effects for different temporal behavior of the SP pulse in a wide domain of two-photon detuning, ranging from slow-photon regime at large values of the latter to temporal oscillations for zero detuning.The remarkable property of our scheme is that this behavior can be tested experimentally along the entire length of photon propagation by using the scale invariance of the SP pulse intensity with respect to the transformation (23). We demonstrated that in our scheme the stored photon can be regenerated into multiple temporal modes with controllable amplitudes, which we have calculated analytically. As an important result, we expect the constructed state of multi-time-bin entangled photon to be applicable to other systems. Further development of this work includes exploring the production of the pairs of time-bin entangled photons with correlated phases, which are highly demanded for creating long-distance quantum channels.

\subsection*{Acknowledgments}%\nonumber

We thank A.Gogyan, H.Jauslin and S.Guerin for helpful discussions. This research has been conducted in the scope of the International Associated Laboratory (CNRS-France and SCS Armenia) IRMAS. We acknowledge additional support from the European Union Seventh Framework Programme Grant No. GA-205025-IPERA, ANSEF Grant PS-opt 3201, and PhD research support program-2012 from SCS of Armenia.

\section{APPENDIX A: SOLUTION FOR FIELD OPERATORS}
In this Appendix we derive the solutions to Eqs.\eqref{9}, \eqref{10} in the general case of arbitrary time dependence of the control field. We introduce an auxiliary function

\begin{equation}
\label{A.1}
\hat{U}(z, \tau)=\int_{0}^zdz'\hat{\sigma}_{12}(z',\tau)\tag{A.1}
\end{equation}

Expressing $\mathcal{\hat{E}}(z,\tau)$ in terms of $\hat{U}(z,\tau)$ from Eq.\eqref{9}
\begin{equation}
\label{A.2}
\hat{\mathcal{E}}(z,\tau)=i2\pi\frac{\omega}{c}\mathcal{N}\mu_{31}\frac{\Omega_c(\tau)}{\Delta}\hat{U}(z, \tau) + \mathcal{\hat{E}}(\tau)\tag{A.2}
\end{equation}
and substituting into Eq.\eqref{10} we find
\begin{equation}
\label{A.3}
\frac{\partial^2\hat{U}}{\partial z\partial\tau}=-i\delta_{tot}(\tau)\frac{\partial\hat{U}}{\partial z}-q\frac{|\Omega_c(\tau)|^2}{\Delta^2}\hat{U}+\hat{F}(\tau) \tag{A.3}
\end{equation}
where
\begin{equation}
\label{A.4}
\hat{F}(\tau)=i\frac{\mu_{31}}{\hbar}\frac{\Omega_c^*(\tau)}{\Delta}\mathcal{\hat{E}}(\tau)\tag{A.4}
\end{equation}

The solution to Eq.(A.3) is found subjected to initial and boundary conditions $U(z, -\infty)=0$ and $U(0,\tau )=0$ and has the form (for the first time a similar solution, but for the Stokes gain, has been found in \cite{41})

\begin{equation}
\begin{split}
\label{A.5}
\hat{U}(z,\tau)=\\
=2z\int_{-\infty}^{\tau}d\tau'\hat{F}(\tau')\text{exp}[-i\int_{\tau'}^{\tau}\delta_{tot}(x)dx]\frac{J_1(\psi(\tau, \tau'))}{\psi(\tau, \tau')}
\end{split}
\tag{A.5}
\end{equation}

where $J_i(\psi)$ is the $i-$th order Bessel function of argument

\begin{equation}
\label{A.6}
\psi(\tau,\tau')=2\sqrt{qz\int_{\tau'}^{\tau}dx\frac{|\Omega_c(x)|^2}{\Delta^2}}\tag{A.6}
\end{equation}

Then, from Eq.(A.2), with  $\Omega_c(\tau)=\Omega_0(\tau)\text{exp}[i\varphi(\tau)]$, we have

\begin{equation}
\begin{split}
\label{A.7}
\mathcal{\hat{E}}(z,\tau)=\hat{\mathcal{E}}(\tau)-2qz\frac{\Omega_0(\tau)}{\Delta^2}\int_{-\infty}^{\tau}d\tau'\hat{\mathcal{E}}(\tau')\Omega_0(\tau')\\
\frac{J_1(\psi(\tau,\tau'))}{\psi(\tau,\tau')}\text{exp}[i(\varphi(\tau)-\varphi(\tau'))-i\int_{\tau'}^{\tau}\delta_{tot}(x)dx]
\end{split}
\tag{A.7}
\end{equation}

\section{APPENDIX B: CALCULATION OF THE PULSE AREA}

To get the solution \eqref{18} for the pulse area we integrate Eq.\eqref{14} over time that yields to the following equation for the pulse area  $\theta(z,\tau)=g\int_{-\infty}^{\tau}\Phi(z, \tau')d\tau'$

\begin{equation}
\begin{split}
\label{33}
\theta(z,\tau)=\theta(0,\tau)-2qz\frac{\Omega_0^2}{\Delta^2}\int_{-\infty}^{\tau}d\tau'\\
\int_{\infty}^{\tau'}dxf(x)e^{-\Gamma(\tau'-x)}\frac{J_1(\psi_0(\tau',x))}{\psi_0(\tau',x)}
\end{split}
\tag{B.1}
\end{equation}

Since $f(t)=d\theta (0,t)/gdt$, the second term of Eq.(B.1) is calculated by the integration by parts leading to

\begin{equation}
\begin{split}
\label{34}
\theta(z,\tau)=\theta(0,\tau)-\\
2qz\frac{\Omega_0^2}{\Delta^2}\int_{-\infty}^{\tau}d\tau'\theta(0,\tau')e^{-\Gamma(\tau-\tau')}\frac{J_1(\psi_0(\tau,\tau'))}{\psi_0(\tau,\tau')}
\end{split}
\tag{B.2}
\end{equation}

Because  $\theta(0, \tau) $ depends on time very weakly it can be taken out from the integral that reduces this equation to the simple form

\begin{equation}
\begin{split}
\label{35}
\theta(z,\tau)=\theta(0,\tau)[1-2qz\frac{\Omega_0^2}{\Delta^2}\int_{-\infty}^{\tau}d\tau' \\e^{-\Gamma(\tau-\tau')}\frac{J_1(\psi_0(\tau,\tau'))}{\psi_0(\tau,\tau')}]
\end{split}
\tag{B.3}
\end{equation}

Here the second term is a known function $1-\text{exp}(-qz\Omega_0^2/\Gamma\Delta^2)$, so that proceeding to the limit $\tau\rightarrow\infty$, we eventually obtain the equation \eqref{18}.

Finally, we derive the relation \eqref{21} from the equation for $\rho_{22}(z,\tau)$

\begin{equation}
\begin{split}
\label{36}
\dot\rho_{22}=2Im\langle\hat{G}^\dag\hat{\sigma}_{12}\rangle=2Im\langle \hat{G}^\dag(z,\tau)\\
i\int_{-\infty}^{\tau}\hat{G}(z,\tau')d\tau'\rangle=2\frac{\Omega_0^2}{\Delta^2}\theta(z,\tau)d\theta(z,\tau)/d\tau
\end{split}
\tag{B.4}
\end{equation}

therefore

\begin{equation}
\rho_{22}(z,\tau)=\frac{\Omega_0^2}{\Delta^2}\theta^2(z,\tau)\tag{B.5}
\end{equation}
which in the limit of $\tau\rightarrow\infty$ coincides with Eq. \eqref{21}.

\end{document}